# Generation of Pure Spin Current with Insulating Antiferromagnetic Materials

Yingwei Chen[1], Junyi Ji[1,2,*], Liangliang Hong[1], Xiangang Wan[3,4,5,6], Hongjun Xiang[1,*]

[1]Key Laboratory of Computational Physical Sciences (Ministry of Education), Institute of Computational Physical Sciences, State Key Laboratory of Surface Physics, and Department of Physics, Fudan University, Shanghai 200433, China

[2]Beijing National Laboratory for Condensed Matter Physics and Institute of Physics, Chinese Academy of Sciences, Beijing 100190, China

[3]National Laboratory of Solid State Microstructures and School of Physics, Nanjing University, Nanjing 210093, China

[4]Collaborative Innovation Center of Advanced Microstructures, Nanjing University, Nanjing 210093, China

[5]Hefei National Laboratory, Hefei 230088, China

[6]Jiangsu Physical Science Research Center, Nanjing 210093, China

Email: jyji@iphy.ac.cn, hxiang@fudan.edu.cn

## Abstract

The generation of pure spin currents is critical for low-dissipation spintronic applications, yet existing methods relying on spin-orbit coupling or ferromagnetic interfaces face challenges in material compatibility and operational robustness. We propose a paradigm-shifting approach to generate symmetry-protected pure spin currents by applying mechanical stress on insulating antiferromagnetic materials, i.e., the pure piezospintronic effect. We first classify magnetic point groups enabling pure piezospintronic effects. A novel first-principles method is developed to compute the spin dipole moments and coefficients of the piezospintronic effect. Integrating these methodologies with high-throughput screening, we identify FeOOH, $Cr_2O_3$ and NaMnX (X=As, Bi, P, Sb) with significant pure piezospintronic effects. Interestingly, we reveal that the ionic displacement contribution dominates the piezospintronic effect, in contrast to the piezoelectric effect. Our study not only provides first-principles approach for investigating spin dipole moment related phenomena (e.g., ferrotoroidicity, fractional quantum spin dipole moment, piezospintronics), but also provide promising piezospintronic materials for experimental verification and industrial applications.

**Main Text**

*Introduction*-Spintronics, as a frontier in condensed matter physics, holds transformative potential for next-generation electronics by harnessing spin currents to circumvent the limitations of charge-based devices [1,2]. The generation of pure spin currents (unaccompanied by net charge flow) is particularly crucial for achieving ultra-low-power operation, eliminating Ohmic losses, and enabling non-volatile information processing. Conventional approaches relying on spin-orbit coupling (SOC) mechanisms—such as the spin Hall effect [3,4] or the Rashba-Edelstein effect [5] suffer from inherent drawbacks [6-9]: (1) Inefficient charge-to-spin conversion due to spin-flip scattering and limited spin diffusion lengths; (2) Mandatory involvement of charge currents, which induces parasitic Joule heating and restricts scalability; (3) Material-specific constraints (e.g., requiring heavy metals like Pt or complex heterostructures).

To discover more effective ways to generate spin currents, we can draw inspiration from charge-current $\boldsymbol{J}^e$ generation in insulators via time-varying polarization as: $\boldsymbol{J}^e = \frac{d\boldsymbol{P}}{dt}$. The polarization $\boldsymbol{P}$ consists of ionic part $\boldsymbol{P}^{\text{ion}}$ and electronic part $\boldsymbol{P}^e = \frac{1}{V_{\text{cell}}} e \langle \hat{\boldsymbol{r}} \rangle_e$ [10], where $V_{\text{cell}}$ is the volume of unit cell and $\langle \hat{\boldsymbol{r}} \rangle_e = \langle \psi_e | \hat{\boldsymbol{r}} | \psi_e \rangle$ is the expectation value of position operator $\hat{\boldsymbol{r}}$ with electronic wavefunction $\psi_e$ in the unit cell. In spintronics, we can similarly define the spin current $J^s_{i;j}$ and spin dipole moment $P^s_{i;j}$ as [11,12]:

$$J^s_{i;j} = \frac{dP^s_{i;j}}{dt}, \qquad P^s_{i;j} = \frac{1}{V_{\text{cell}}} \langle \hat{S}_i \hat{r}_j \rangle_e, \qquad (1)$$

where $\hat{S}_i$ is the spin operator, $i$ and $j$ label the spin and position components, respectively. Therefore, changes in the spin dipole moment can lead to spin currents. For example, considering the spin dipole moment changes caused by mechanical deformations, akin to the piezoelectric effect, the piezospintronic effect [12,13] was proposed:

$$\Delta P_{i;j}^S = \sum_{kl} \lambda_{i;jkl} \eta_{kl}, \tag{2}$$

with $\lambda$ being the piezospintronic pseudo tensor, and $\eta_{kl}$ denoting the strain tensor, as shown in Fig. 1.

It was showed that antiferromagnetic crystals preserving $\mathcal{PT}$ symmetry may generate pure spin currents (without electronic current) under strain [12,13], where $\mathcal{P}$ and $\mathcal{T}$ represent space inversion and time reversal symmetries, respectively. This is rigorously demonstrated through symmetry operations on spin ($\boldsymbol{s}$) and position ($\boldsymbol{r}$) as $\mathcal{T}: \boldsymbol{s}, \boldsymbol{r} \to -\boldsymbol{s}, \boldsymbol{r}$ and $\mathcal{P}: \boldsymbol{s}, \boldsymbol{r} \to \boldsymbol{s}, -\boldsymbol{r}$. Consequently, the spin dipole moment ($P^S \propto \boldsymbol{s} \otimes \boldsymbol{r}$) in magnetic systems transforms as: $P^S \to -P^S$ (under $\mathcal{P}$ or $\mathcal{T}$ alone) and $P^S \to P^S$ (under $\mathcal{PT}$). On the other hand, for the electric dipole moment as $e\boldsymbol{r}$ changes sign under $\mathcal{P}$ or $\mathcal{PT}$ symmetry. Thus, for magnetic crystals preserving $\mathcal{PT}$ symmetry may have pure piezospintronic effect. This enables insulating antiferromagnetic (AFM) materials to produce pure spin currents without involving any net electronic current. Furthermore, controlling the magnetic moment of insulating AFM materials could also enable spin-current generation, a crucial mechanism in AFM spintronics, which can be achieved through magnetic or electric fields, charge currents, or ultrafast THz pumping [14,15]. Recently, it was proposed that altermagnetic systems [16,17] can generate spin currents through their intrinsic spin-split electronic bands and nonrelativistic symmetry-breaking mechanisms. However, achieving efficient spin current generation in metallic altermagnets (e.g., $RuO_2$, CrSb) requires finite charge currents, inevitably inducing Joule heating due to electron scattering.

However, the concepts of spin dipole moment and piezospintronic effect were suggested through a simple tight-binding analysis [13,18] that only considers a pure electronic clamped-ion contribution and whether the piezospintronic effect can be significant is not known. Moreover, methods for calculating the spin dipole moment and piezospintronic effect from first-principles are not available, and promising materials with a significant piezospintronic effect have not been identified.

In this Letter, we perform symmetry analysis to identify magnetic point groups enabling spin dipole moments and pure piezospintronic effects. We further develop a density functional theory (DFT) based method to determine these two quantities in insulating magnetic materials. A high-throughput first-principles screening workflow is then performed to discover promising materials with significant piezospintronic effect capable of generating pure spin current.

*Symmetry analysis*-Through group theory, we can identify the magnetic point groups that allow non-zero spin dipole moment and pure piezospintronic effect, which is summarized in Table 1 and details could be found in Section 1 of Supplementary Materials (SM) [19]. Notably, 21 magnetic point groups can host a pure non-zero spin dipole moment without requiring $\mathcal{PT}$ symmetry. Moreover, for pure piezospintronic effect, most of the allowed groups contain the $\mathcal{PT}$ symmetry, with the exceptions of the 432.1 and 4'32'. Thus, $\mathcal{PT}$ symmetry is neither sufficient nor necessary for these two quantities. For systems with negligible spin-orbit coupling, spin-space decoupling introduces additional symmetry constraints: a magnetic group may allow a non-zero spin dipole moment, but the corresponding spin group [20-23] may forbid it.

Notably, analogous to electric polarization, only differences in spin dipole moments between two distinct configurations are physically measurable. Interestingly, there may be non-zero difference in the spin dipole moment between two states where the native spin dipole moments are forbidden according to the usual symmetry analysis. Specifically, by displacing a magnetic atom $k$ from one high symmetric Wyckoff position to an equivalent one, a non-zero spin dipole moment can be generated, as $\Delta P^s_{i;j} \approx \frac{1}{V_{\text{cell}}} \sum_k m_i^{(k)} \Delta \tau_j^{(k)}$, where $m_i^{(k)}$ is the magnetic moment in the $i$ direction of atom $k$ and $\Delta \boldsymbol{\tau}^{(k)}$ is its displacement. This is similar to the concept of fractional quantum ferroelectricity [24,25] (see Fig. 1) that can display in a non-polar crystal.

*First-principles methods*-Similar to the derivation of electric polarization and orbital magnetism [10,26] in DFT, the spin dipole moment density could be expressed using Wannier functions as (in Hartree atomic units with $\hbar = e = m_{\text{e}} = 1$):

$$P_{i;j}^{s} = \frac{1}{V_{\text{cell}}} \sum_{n}^{\text{occ}} \langle w_{0n} | \hat{S}_i \hat{r}_j | w_{0n} \rangle = \frac{i}{(2\pi)^3} \sum_{n}^{\text{occ}} \int_{\text{BZ}} d^3 k \langle u_{nk} | \left(\frac{1}{2}\hat{\sigma}_i\right) | \partial_{k_j} u_{nk} \rangle \quad (3)$$

where $|w_{Rn}\rangle$ is real-space Wannier function of cell $R$, and $|u_{nk}\rangle = e^{-ikr}|\psi_{nk}\rangle$ is the periodic part of the Bloch function, $n$ labels the band component and the summation is over all occupied states. 'BZ' represents the first Brillouin zone and $\hat{\sigma}_i$ is the Pauli matrix. Details of the derivation can be found in Section 2 of SM.

For collinear magnetism without considering SOC, assuming that the spin directions are aligned along the z-axis, the spin dipole moment could be simplified through the electric polarization $P^e$, as:

$$P_{z;j}^{S} = \frac{1}{2}\left(P_{\uparrow,j}^{e} - P_{\downarrow,j}^{e}\right), \quad (4)$$

where $P_{\uparrow,j}^{e}$ and $P_{\downarrow,j}^{e}$ are the spin-up and spin-down contribution of $P_j^e$, respectively.

However, for noncollinear cases or when SOC is considered, a generalized approach becomes essential. The similar quantity has already been calculated using Wannier functions in spin Hall effect [27,28] or magnetoelectric monopole density [29]. In their methods, the unitary transformation matrix $V_{k,mn}$ is obtained through the maximally localized Wannier functions [30], satisfying:

$$\left|u_{nk}^{(w)}\right\rangle = \sum_{R} e^{-ik\cdot(r-R)} |w_{Rn}\rangle, \quad \left|u_{nk}^{(w)}\right\rangle = \sum_{m} \left|u_{mk}^{(0)}\right\rangle V_{k,mn}, \quad (5)$$

where $\left|u_{mk}^{(0)}\right\rangle$ is the periodic part of the Bloch function obtained through DFT. The spin dipole moment could thus be calculated as:

$$P_{i;j}^{s} = i\frac{1}{V_{\text{cell}} N_k} \sum_{n}^{\text{occ}} \sum_{k} \sum_{b_j} w_{b_j} b_j \left[V_k^+ S_k^{(0)} M_{k,b_j}^{(0)} V_{k+b_j} - V_k^+ S_k^{(0)} V_k\right]_{nn} \quad (6)$$

where $S_{k,mn}^{(0)} = \langle u_{mk}^{(0)} | \hat{\sigma}_i | u_{nk}^{(0)} \rangle$, $M_{k,b,mn}^{(0)} = \langle u_{mk}^{(0)} | u_{n,k+b}^{(0)} \rangle$, $N_k$ is the total number of $k$ points in the DFT calculations and $b$ is the vector connecting a $k$-point to its near neighbors and together with its weight $w_b$. Details of the method could be found in Section 3 of SM. Unfortunately, this method proves inaccurate in practice because the quantity in Eq. (6) is gauge-dependent [31,32], unlike Eq. (4). By applying a random phase to the wavefunction $u_{nk}$, Eq. (6) may result in a different outcome. To eliminate this problem, we adopt the same approach used in Wannier centers [30] and correct it as:

$$P_{i;j}^S = \frac{-1}{V_{\text{cell}} N_k} \sum_n^{\text{occ}} \sum_k \sum_{b_j} w_{b_j} b_j \, S_{k,nn}^{(w)} \cdot \text{Im} \ln \left[ \frac{\left[ S_k^{(w)} M_{k,b_j}^{(w)} \right]_{nn}}{S_{k,nn}^{(w)}} \right], \quad (7)$$

where $S_k^{(w)} = V_k^+ S_k^{(0)} V_k$ and $M_{k,b}^{(w)} = V_k^+ M_{k,b}^{(0)} V_{k+b}$. Details of the derivation and other practical corrections could be found in Section 4 of SM. These corrections ensure that the results are invariant to phase changes in isolated bands, i.e., $\left| u_{nk}^{(0)} \right\rangle \to e^{i\phi_n(k)} \left| u_{nk}^{(0)} \right\rangle$ where $\phi_n(k)$ is a real function.

We could calculate the piezospintronic coefficients using the finite difference method, as illustrated by the expression:

$$\lambda_{i;jkl}^{\text{improper}} = \frac{1}{\delta} \left[ P_{i;j}^S(\eta_{kl} = \delta) - P_{i;j}^S(\eta_{kl} = 0) \right], \quad (8)$$

where $\delta$ is a small number. Also, similar to piezoelectric effect [33], we could define proper piezospintronic coefficients that correspond to the experimental measurement, which can be computed as:

$$\lambda_{i;jkl}^{\text{proper}} = \lambda_{i;jkl}^{\text{improper}} + \delta_{kl} P_{i;j}^S - \delta_{jk} P_{i;l}^S. \quad (9)$$

Additionally, the piezospintronic effect can be attributed to two main components: clamped-ion and internal-strain contribution. Clamped-ion contribution is calculated by fixing the atomic fractional coordinates under strain. Internal-strain refers to the internal distortion of the nuclear coordinates at fixed strain, which could be simply approximated by:

$$\Delta P_{i;j}^{S,\text{ion}} \approx \frac{1}{V_{\text{cell}}} \sum_{q,k} m_{\text{eff};i,j,q}^{(k)} \Delta \tau_q^{(k)}, \quad (10)$$

where $\Delta \tau_q^{(k)}$ is the displacement of atom $k$ along $q$ direction and $m_{\text{eff};i,j,q}^{(k)} = V_{\text{cell}} \frac{dP_{i;j}^S}{d\tau_q^{(k)}}$ is the effective magnetic moment as a third-rank tensor, which could also be roughly approximated as the atomic magnetic moment $m_i^{(k)}$. Additionally, $\Delta \tau_q^{(k)}$ could be calculated using the force-constant matrix $K_{mn}^{(k)}$ and the force-response internal-strain tensor $\Lambda_{m;kl}^{(k)}$, which are defined by [34]:

$$K_{mn}^{(k)} = V_{\text{cell}} \frac{\partial^2 E}{\partial \tau_m^{(k)} \partial \tau_n^{(k)}}\bigg|_\eta , \quad \Lambda_{m;kl}^{(k)} = -V_{\text{cell}} \frac{\partial^2 E}{\partial \tau_m^{(k)} \partial \eta_{kl}} , \quad (11)$$

where $E$ presents the energy per undeformed unit cell volume. Thus, the internal-strain piezospintronic coefficient could be approximated by

$$\lambda_{i;jkl}^{\text{ion}} \approx \frac{1}{V_{\text{cell}}} \sum_{k,q,n} m_{\text{eff};i,j,q}^{(k)} \left(K^{(k),-1}\right)_{qn} \Lambda_{n;kl}^{(k)} . \quad (12)$$

*Application to MnPS$_3$*-2D MnPS$_3$ has been suggested as an candidate material [13] for pure piezospintronic effect. Its Néel antiferromagnetic ground state, confirmed experimentally [35] and via DFT [36], remains stable under strain [36]. The layer group of 2D MnPS$_3$ is P$\bar{3}$1m, whose magnetic structure is illustrated in Fig. 2(a). The usual symmetry analysis suggests zero $P_{z;y}^s$ without strain, but finite piezospintronic coefficient $\lambda_{z;yyy}$. For simplicity, we consider the phase of spin dipole moment $\phi_{z;y}^s = \frac{V_{\text{cell}}}{b} P_{z;y}^s$ in the following ($b$ is the length of the lattice vector $\boldsymbol{b}$ along the $y$ direction). For different strain $\eta_{yy}$, the calculated phase $\phi_{z;y}^s$ is shown in Fig. 2(b). Our calculations show a piezospintronic coefficient of $\lambda_{z;yyy}^{\text{proper}} = -0.029\,\mu_B\text{Å}^{-1}$ (using cell area for 2D systems). The previous Wannier method as Eq. (6) exhibits significant deviations from Eq. (4) results even with a denser $\boldsymbol{k}$-point grid. In contrast, the corrected Wannier approach as Eq. (7) achieves quantitative agreement, confirming methodological consistency.

Notably, $P_{z;y}^s$ at zero strain is computed to be $0.11\mu_B\text{Å}^{-1}$ ($V_{\text{cell}}$ is replaced with the area in Eq. (1) for 2D system), contradicting the magnetic point group symmetry analysis. In fact, this exception is similar to the mechanism of fractional quantum ferroelectricity. A comprehensive discussion of the fractional quantum property in 2D MnPS$_3$ is presented in Section 5 of SM.

We could also compare the spin dipole moment with versus without SOC in Fig.2 (b), revealing negligible SOC influence on spin dipole moments or piezospintronic responses, contrasting sharply with spin Hall effects. This is because SOC does not significantly change the magnitude of the magnetic moments governing spin dipole formation. Details of the calculation and other examples could be found in Section 6 and Section 7 of SM.

*High-throughput screening*-To identify materials with a significant piezospintronic effect for generating pure spin current, we conduct a high-throughput screening process, as illustrated in Fig. 3. Our search begins with 2,138 experimentally observed magnetic materials from the MAGNDATA database [37,38]. We first exclude ferromagnetic materials by calculating the total magnetization. We then utilize symmetry analysis (implemented in the Property Analysis and Simulation Package for materials (PASP) [39]) to predict non-zero tensors of piezoelectric coefficients and piezospintronic coefficients, and thus eliminate piezoelectric and non-piezospintronic systems. Next, as the calculation method requires insulating materials, we search the band gap data of these candidates in the Materials Project [40], ensuring the same chemical composition, magnetic state, and space group to identify insulators. Finally, for practical device applications, we exclude materials with a Néel temperature below 77K.

This process results in 85 candidate materials exhibiting the piezospintronic effect, as listed in Table S2 of SM. Examining these materials more closely, we find that they all exhibit $\mathcal{PT}$ symmetry. This is because only BaCuTe$_2$O$_6$ [41] and SrCuTe$_2$O$_6$ [42] in the database belongs to the magnetic point group of 432.1 or 4'32' and their transition temperatures are about 5K. Symmetry analysis via magnetic point group and spin point group [20] reveals that 26 collinear materials exhibit piezospintronic effects induced by SOC. Despite SOC's role, its low magnitude in these materials leads to minimal piezospintronic responses.

Further DFT calculations are then conducted to verify the remaining 41 collinear magnetic materials where the effect is not caused by SOC and 18 noncollinear magnetic materials. After high throughput calculations, materials with a large piezospintronic effect are presented in Table 2.

Among the listed materials, the band gap of AgRuO$_3$ is about 80 meV both from DFT [43] and experiment [44], which limits its utility in devices. Considering the fact that GeMnO$_3$ possesses a Néel temperature substantially below those of other candidates, hereafter we will focus on FeOOH, Cr$_2$O$_3$ and NaMnX (X=As, Bi, P, Sb)

The FeOOH here refers to α-FeOOH (Pnma' magnetic space group), as illustrated in Fig. 4(a). This material exhibits a high Néel temperature exceeding 340K and is one of the most common oxyhydroxides on Earth [45,46]. Both experimental studies and first-principles calculations indicate that the band gap of the system is about 2 eV [45], which makes it promising for industrial devices. As shown in Fig. 4(a), applying strain causes the magnetic ions to displace, with ions of the same spin direction moving in one direction, resulting in the piezospintronic effect. The piezospintronic response in Fig. 4(b) is characterized by $\Delta\phi = \phi(\eta) - \phi(\eta = 0)$, where $\phi(\eta)$ and $\phi(\eta = 0)$ denote the phases of the spin dipole moment under applied strain $\eta$ and at equilibrium, respectively. Additionally, Fig. 4(b) displays the internal-strain term, calculated by Eq. (10) along with effective magnetic moments determined through finite differences. The coefficient of the internal-strain is about $\lambda^{\text{ion}}_{3;211} = 0.046\ \mu_B \cdot \text{Å}^{-2}$, which is close to $\lambda^{\text{tot}}_{3;211} = 0.042\ \mu_B \cdot \text{Å}^{-2}$. Unlike the piezoelectric effect [34], the results show that the internal-strain contribution dominates the total effect. Also, we could estimate the magnitude of the spin current generated through piezospintronic effect in the FeOOH. Considering the formalism as $J^S_{i;j} = \frac{\Delta P^S_{i;j}}{\Delta t} = \frac{\Delta P^S_{i;j}}{\Delta \eta_{kl}} \frac{\Delta \eta_{kl}}{\Delta t} = \lambda_{i;jkl} \frac{\Delta \eta_{kl}}{\Delta t}$, the spin current density scales with both the applied strain magnitude $\Delta \eta_{kl}$ and its temporal rate $\Delta \eta_{kl}/\Delta t$. For quantification, we adopt experimental parameters from a piezoelectric study [47] as $\Delta \eta_{kl} = 0.5\%$ and $\Delta t \approx 20$ ms, yielding the spin current as $J^S = 1.17 \times 10^{11}\ (\hbar/2e)\ \text{A} \cdot \text{m}^{-2}$, which is close to the magnitude of the spin current density generated by spin Hall effect in Pt [48,49]. Further discussion of the spin current generated through piezospintronic effect could be found in SM Section 9.

Chromia (Cr$_2$O$_3$) is an established collinear magnetoelectric antiferromagnet [50], showing both a high Néel temperature and wide band gap. Our study reveals that Cr$_2$O$_3$ also demonstrates exceptional piezospintronic performance. Using a DFT-parameterized model Hamiltonian [51] detailed in Section 11 of SM, we simulated the temperature dependence of its spin dipole moment. The results show the spin dipole moment (and thus the piezospintronic coefficient) follow $P_{i;j}^S(T) \approx P_{i;j}^S(T=0) \times |\bar{\boldsymbol{m}}(T)|$, where $|\bar{\boldsymbol{m}}(T)|$ is the sublattice magnetization.

NaMnX compounds exhibit both high Néel temperatures above 293K and large band gaps of about 2 eV [52]. Using NaMnAs as an example, its magnetic structure is illustrated in Fig. S7(a) of SM, which belongs to the P4'/n'm'm magnetic space group. As shown in Fig. S7(b), the clamped-ion contribution is negligible. Also, The Mn ion displacement $\Delta\tau_y^{(Mn,\uparrow)}$ agrees with the spin dipole moment behavior. The result shows that the spin dipole moment exhibits the same behavior as the lattice distortion under applied strain in this system. We also compared materials exhibiting a large piezospintronic effect as NaMnAs with those showing a smaller effect as bulk MnPS$_3$ ($\lambda_{3;222} = -0.0065\ \mu_B \cdot \text{Å}^{-2}$). As depicted in Fig. S7(b), both systems exhibit negligible clamped-ion effects. In bulk MnPS$_3$, reduced displacement of Mn atoms and doubled unit cell volume compared to NaMnAs halve its internal-strain contribution. This aligns with Eq. (12), which quantifies enhanced piezospintronic performance through four factors: high magnetic atomic density, large magnetic moments, strong force-response internal-strain tensor, and low force-constants such as soft-mode systems. These criteria establish a roadmap for designing strain-engineered spintronic materials.

*Summary*-We perform group theory analysis to show that $\mathcal{PT}$ symmetry is neither sufficient nor necessary for pure piezospintronic effects. We developed for the first time a first-principles-based computational framework to compute spin dipole moments, piezospintronic coefficients, and magnetic toroidal moments in ferrotoroidicity. High-throughput screening of the MAGNDATA database identified FeOOH, Cr$_2$O$_3$ and NaMnX (X=As,Bi,P,Sb) as promising candidates in which internal-strain contributions is discovered to be responsible for the strong piezospintronic effect. Our work advances understanding of piezospintronic effect and provides experimentally testable materials for industrial applications.

## Acknowledgement

We thank Dr. Lei Shen for the discussion of spin currents induced by spin Hall effects and piezospintronic effect. We acknowledge financial support from NSFC (No. 12188101), the National Key R&D Program of China (No. 2022YFA1402901), Shanghai Science and Technology Program (No. 23JC1400900), the Guangdong Major Project of the Basic and Applied Basic Research (Future functional materials under extreme conditions--2021B0301030005), Shanghai Pilot Program for Basic Research—FuDan University 21TQ1400100 (23TQ017), the robotic AI-Scientist platform of Chinese Academy of Science, and New Cornerstone Science Foundation. J. J. also acknowledges the support from China National Postdoctoral Program for Innovative Talents (BX20230408).

| | Electric | Spin |
|---|---|---|
| **Dipole Moment** | 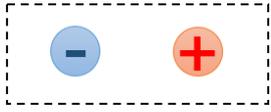 $P_x^e = \sum_i q_i \cdot x_i$ | 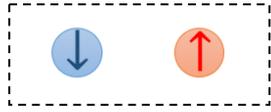 $P_{z;x}^s = \sum_i m_{z;i} \cdot x_i$ |
| **Current** | $J_j^e = \dfrac{\mathrm{d}P_j^e}{\mathrm{d}t}$ | $J_{\sigma;j}^s = \dfrac{\mathrm{d}P_{\sigma;j}^s}{\mathrm{d}t}$ |
| **Piezo Effect** | 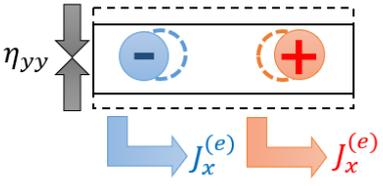 | 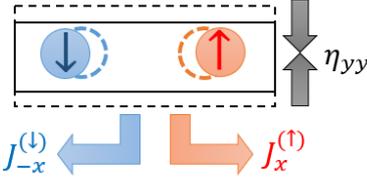 |
| **Fractional Quantum** | 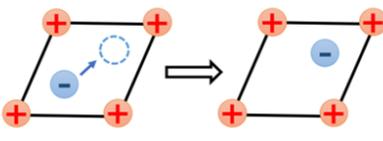 | 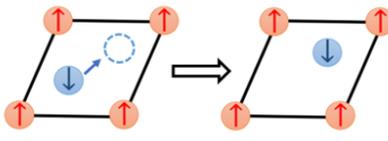 |

Figure 1. The concepts of spin dipole moment, piezospintronic effect and fractional quantum ferroelectricity. In the toy model, the unit cell consists of two points $i$ with opposite charges $q_i$ or opposite magnetic moments $m_{z;i}$. When a strain $\eta_{yy}$ is applied, the system generates an electric current $J_x^{(e)}$ or a spin current $J_x^{(s)}$.

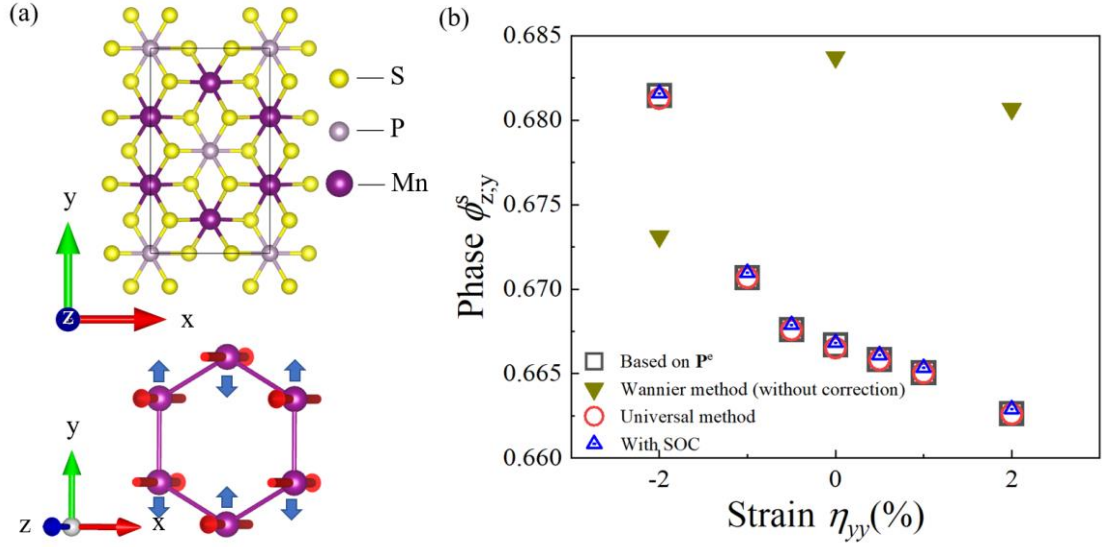

Figure 2. (a) Néel antiferromagnetic structure of 2D MnPS$_3$. The blue arrows beside the Mn ions indicate the displacement of the ions along the y-axis under $\eta_{yy} > 0$. (b) Strain-dependent spin dipole moment phase $\phi^s_{z;y}$ calculated via: Eq. (4) (square points); Eq. (6) (inverted triangle points, some out-of-range), and Eq. (7) (circle points). Triangle points account for results considering spin-orbit coupling.

Table 1. Magnetic point groups in UNI notation [53] hosting non-zero spin dipole moment, pure spin dipole moment, and pure piezospintronic effect. Blue entries denote groups lacking $\mathcal{PT}$ symmetry.

| Crystal system | Spin dipole moment | Spin dipole moment without electric dipole moment | Piezospintronic effect without piezoelectric effect |
|---|---|---|---|
| **Triclinic** | 1.1, -1' | -1' | -1' |
| **Monoclinic** | 2.1, 2', m.1, m', 2'/m, 2/m' | 2'/m, 2/m' | 2'/m, 2/m' |
| **Orthorhombic** | 222.1, 2'2'2', mm2.1, m'm2', m'm'2, m'mm, m'm'm' | 222.1, 2'2'2', m'mm, m'm'm' | m'mm, m'm'm' |
| **Tetragonal** | 4.1, 4', -4.1, -4', 4/m',4'/m', 422.1, 4'22', 42'2', 4mm.1, 4'm'm, 4m'm', -42m.1, -4'2'm, -4'2m', -42'm', 4/m'mm, 4/m'm'm, 4/m'm'm' | -4.1, -4', 4/m',4'/m', 422.1, 4'22', 42'2', -42m.1, -4'2'm, -4'2m', -42'm', 4/m'mm, 4/m'm'm, 4/m'm'm' | 4/m', 4'/m', 4/m'mm, 4/m'm'm, 4/m'm'm' |
| **Trigonal** | 3.1, -3', 32.1, 32', 3m.1, 3m', -3'm, -3'm' | -3', 32.1, 32', -3'm, -3'm' | -3', -3'm, -3'm' |
| **Hexagonal** | 6.1, -6', 6/m', 622.1, 62'2', 6mm.1, 6m'm', -6'm'2, -6'm2', 6/m'mm, 6/m'm'm | -6', 6/m', 622.1, 62'2', -6'm'2, -6'm2', 6/m'mm, 6/m'm'm | 6'/m, 6/m'. 6/m'mm, 6'/mmm', 6/m'm'm' |
| **Cubic** | 23.1, m'-3', 432.1, -4'3'm, m'-3m' | 23.1, m'-3', 432.1, -4'3'm, m'-3m' | m'-3, 432.1, 4'32', m'-3m, m'-3m' |

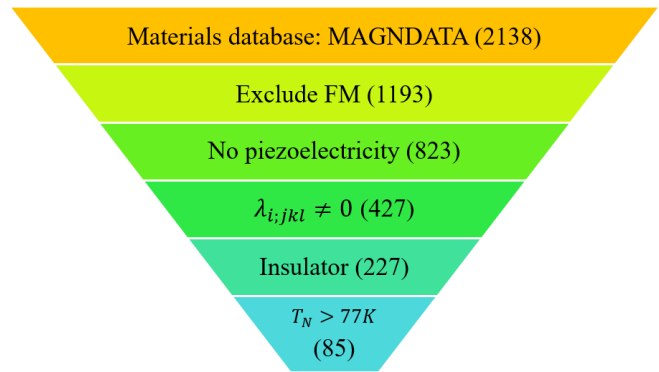

Figure 3. The high-throughput screening workflow. The number of materials remaining at each stage of the process is indicated in parentheses.

Table 2. Materials exhibiting a significant piezospintronic effect ( $|\lambda| > 0.019\ \mu_B \text{Å}^{-2}$). The first row of the table lists the material label from MAGNDATA, number of sites, formula, Néel temperature recorded in MAGNDATA, band gap $E_g$, the piezospintronic effect coefficient and its corresponding value.

| label | sites | formula | $T_N$(K) | $E_g$ (eV) | coefficient | value of λ ($\mu_B\ \text{Å}^{-2}$) |
|---|---|---|---|---|---|---|
| 0.629 | 6 | NaMnAs | >293 | 1.65 | $\lambda_{3;322} = -\lambda_{3;311}$ | 0.0227 |
| 0.635 | 6 | NaMnBi | >293 | 1.00 | $\lambda_{3;322} = -\lambda_{3;311}$ | 0.0196 |
| 0.626 | 6 | NaMnP | >293 | 0.88 | $\lambda_{3;322} = -\lambda_{3;311}$ | 0.0192 |
| 0.631 | 6 | NaMnSb | >293 | 1.32 | $\lambda_{3;322} = -\lambda_{3;311}$ | 0.0216 |
| 0.110 | 10 | Cr$_2$O$_3$ | 307.6 | 1.94 | $\lambda_{3;333}$ | -0.0237 |
| 0.125 | 10 | GeMnO$_3$ | 120 | 1.14 | $\lambda_{3;322} = \lambda_{3;311}$ | 0.0346 |
| 0.399 | 16 | FeOOH | 340-440 | 1.76 | $\lambda_{3;211}$ | 0.0417 |
|  |  |  |  |  | $\lambda_{3;222}$ | 0.0225 |
| 0.733 | 20 | AgRuO$_3$ | 342 | 0.08 | $\lambda_{3;322} = \lambda_{3;311}$ | 0.0301 |

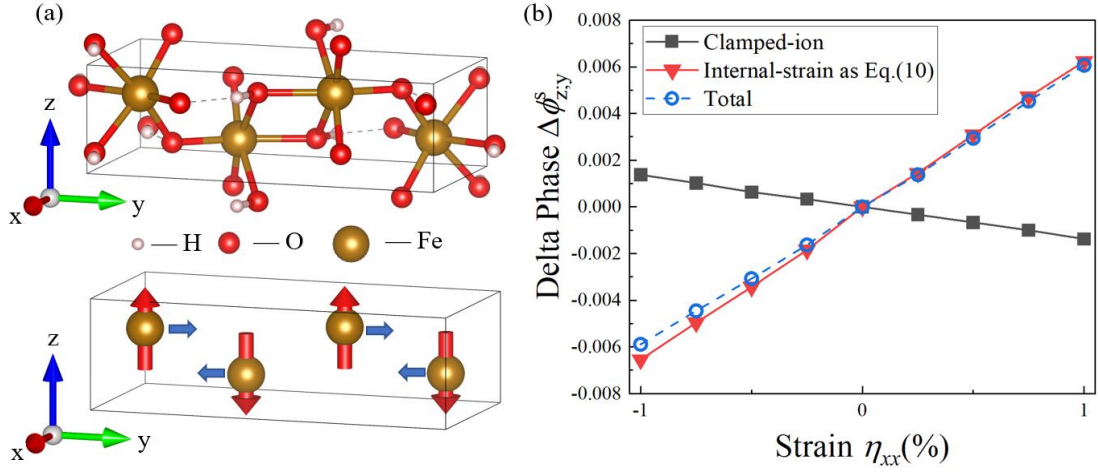

Figure 4. (a) Structure of α-FeOOH. The red arrows on the Fe ions indicate the directions of magnetic moments. The blue arrows beside the Fe ions indicate the displacement of the ions along the y-axis under $\eta_{xx} > 0$. (b) Change in the phase of spin dipole moment $\phi^s_{z;y}$ versus applied strain $\eta_{xx}$ in α-FeOOH, including clamped-ion contribution (square points), internal-strain contribution calculated through Eq. (10) (inverted triangle points) and the total effect (circle points).

# End Matter

**Appendix A: Calculate ferrotoroidicity using spin dipole moment.** Ferrotoroidicity, the fourth form of primary ferroic order, breaks both space and time-inversion symmetry [54]. The toroidization associated with ferrotoroidicity is defined as [29,55]: $\boldsymbol{T}^{\text{cla}} = \frac{1}{2V_{\text{cell}}} \sum_k \boldsymbol{r}^{(k)} \times \boldsymbol{m}^{(k)}$, where $\boldsymbol{r}^{(k)}$ represents the position of the magnetic moments $\boldsymbol{m}^{(k)}$ relative to a chosen point within the unit cell. Classical method approximates $\boldsymbol{r}^{(k)}$ as the position of the $k$-th magnetic atom [55], limiting accuracy for delocalized magnetic moments. Using our framework, we calculated the toroidization in LiCoPO$_4$, yielding $\boldsymbol{T}_z^{\text{cla}} = (P_{y;x}^s - P_{x;y}^s)/2 = 5.98 \times 10^{-3} \mu_B \text{Å}^{-2}$, which closely matches the previously reported value of $\boldsymbol{T}_z^{\text{cla}} = 6.17 \times 10^{-3} \mu_B \text{Å}^{-2}$ using the classical approximation [55]. And our approach could overcome localization constraints inherent to classical models.